\documentclass[fleqn,10pt]{wlscirep}
\usepackage[utf8]{inputenc}
\usepackage[T1]{fontenc}
\usepackage{bm}

\usepackage{dcolumn}
\usepackage{bm}
\usepackage{xcolor}
\usepackage{enumitem}
\usepackage{hyperref}
\usepackage{verbatim}
\usepackage{amsmath,amstext,mathrsfs}
\usepackage[all]{hypcap} 
\usepackage{afterpage}    
\usepackage{marginnote}
\usepackage{makecell}
\usepackage{subfigure}
\usepackage{xfrac}
\usepackage{makecell}
\usepackage{float}

\usepackage{ifxetex,ifluatex}
\usepackage{fixltx2e} 
\usepackage{multirow}

\DeclareFontFamily{OMS}{oasy}{\skewchar\font48 }
\DeclareFontShape{OMS}{oasy}{m}{n}{%
         <-5.5> oasy5     <5.5-6.5> oasy6
      <6.5-7.5> oasy7     <7.5-8.5> oasy8
      <8.5-9.5> oasy9     <9.5->  oasy10
      }{}
\DeclareFontShape{OMS}{oasy}{b}{n}{%
       <-6> oabsy5
      <6-8> oabsy7
      <8->  oabsy10
      }{}
\DeclareSymbolFont{oasy}{OMS}{oasy}{m}{n}
\SetSymbolFont{oasy}{bold}{OMS}{oasy}{b}{n}

\DeclareMathSymbol{\smallleftarrow}     {\mathrel}{oasy}{"20}
\DeclareMathSymbol{\smallrightarrow}    {\mathrel}{oasy}{"21}
\DeclareMathSymbol{\smallleftrightarrow}{\mathrel}{oasy}{"24}

\newcommand*{\hy}{} 
\graphicspath{{./}{Fig/}}

\title{Temporal Properties of the Compressible Magnetohydrodynamic Turbulence}

\author[1,*]{Ka Ho Yuen}
\author[1]{Hui Li}
\author[2,3]{Huirong Yan}
\affil[1]{Theoretical Division, Los Alamos National Laboratory, Los Alamos, NM 87545, USA}
\affil[2]{Deutsches Elektronen-Synchrotron DESY, Germany}
\affil[3]{Institut für Physik \& Astronomie, Universität Potsdam, Germany}
\affil[*]{\it Corresponding Author: kyuen@lanl.gov ,ORCID: 0000-0003-1683-9153}

\begin{abstract}
The temporal property of the compressible magneto-hydrodynamic (MHD) turbulence remains a fundamental unsolved question. 
Recent studies based on the spatial-temporal analysis in the global frame of reference suggest that the majority of fluctuation power in turbulence does not follow any of the MHD wave dispersion relations but has very low temporal frequency with finite wavenumbers.
Here, we demonstrate that the Lorentzian broadening of the dispersion relations of the three MHD modes where the nonlinear effects act like the damping of a harmonic oscillator can explain many salient features of frequency spectra for all MHD modes. 
The low frequency fluctuations are dominated by modes with the low parallel wavenumbers that have been broadened by the nonlinear processes.
The Lorentzian broadening widths of the three MHD modes exhibit scaling relations to the global frame wavenumbers and are intrinsically related to energy cascade of each mode.
Our results provide a new window to investigate the temporal properties of turbulence which offers insights for building a comprehensive understanding of the compressible MHD turbulence.
\end{abstract}
\begin{document}

\flushbottom
\maketitle

\thispagestyle{empty}

\section*{Main}

There has been a long history in studying the temporal properties of MHD turbulence
\cite{K41,1964SvA.....7..566I,1965PhFl....8.1385K}, particularly
the origin of low frequency temporal fluctuations\cite{2021PhPl...28c2306M}. 
These low-frequency fluctuations  
have implications for several problems in both space physics and astrophysics 
\cite{2012ApJ...745...35Z},
including the heating of the solar corona  \cite{1999ApJ...523L..93M},
the low-frequency ``$1/f$ noise" in the solar wind \cite{2007PhRvE..76c6305D,2009PhPl...16f2304D},
the formation and evolution of stars and molecular clouds 
in the interstellar medium \cite{MO07}, 
as well as the propagation and acceleration of cosmic rays 
\cite{1998JGR...103.2085Z,2002PhRvL..89B1102Y,2006ApJ...642..230S}.
Some of the earliest theoretical models came from the ``2D plus slab" model in the
nearly incompressible magnetohydrodynamics 
\cite{1990JGR....9520673M,1992JGR....9717189Z},
suggesting that a perpendicular cascade (i.e., the 2D fluctuations with global frame parallel wavenumber $k_\parallel=0$\footnote{
The notion of $k_\parallel=0$ is defined in simulations that have periodic boundary condition parallel to the global mean field direction. The fluctuation of $k_\parallel=0$ refers to the infinite integration of variations along the global mean field. However, in realistic astrophysical settings, all astrophysical objects are bounded by a certain characteristic length $L$ \cite{2010ApJ...710..853C} which defines the minimal wavenumber $k_{min} = 2\pi/L$, therefore the notion of $k_\parallel=0$ has questionable physical meaning. However, the role of $k_\parallel=0$ fluctuations in the energy transfer of both in weak \cite{1996ApJ...465..845N,2007NJPh....9..307N} and strong \cite{GS95} turbulence should not be neglected. }) 
could generate nearly zero-frequency fluctuations. 
Alternatively, the {\it non-resonant} three-wave interaction in strong Alfv\'enic turbulence 
could generate MHD fluctuations at $k_\parallel \approx 0$ and 
$\omega \approx 0$ \cite{GS95,2000ApJ...539..273C}. 
Different views have fueled the debate whether to treat these low frequency
fluctuations as waves or nonlinear structures\cite{2004RvMP...76.1015Z}. 
Several physical pictures were put forth to interpret the low-frequency fluctuations  
such as those from damped harmonic oscillators \cite{2013PhPl...20g2302H,2014CoPhC.185..578T}, 
sweeping modes\cite{2004RvMP...76.1015Z,2016PhPl...23k2304L}, magnetosonic modes \cite{2017PhPl...24j2314A}
and also transition from weak to strong turbulence \cite{2016PhRvL.116j5002M}. 

One interesting approach is to perform spatio-temporal analysis on simulated turbulence fluctuations
in order to extract their properties
\cite{2007PhRvE..76c6305D,2009PhPl...16f2304D,Sv09,2013PhPl...20g2302H,
2016PhRvL.116j5002M,2016PhPl...23k2304L,2017PhPl...24j2314A,
2019PhPl...26l2301L,2019MNRAS.488..859Y,2020PhRvX..10c1021M} and gain 
unique insights on how turbulence evolves in space and time.
In particular, recent numerical studies \cite{4DFFT,2022ApJ...936..127F} 
have quantified the spatio-temporal distribution of velocity, magnetic field and density variations, showing that they not only deviate from the simple dispersion relations for compressible MHD modes but also hold a dominant fraction in low temporal frequencies, which is also observed from satellite observations\cite{2010JGRA..115.4101N,2011GeoRL..38.5101N,2023ApJ...944...98Z,2023arXiv230106709Z}. 

In this paper, we explain quantitatively how the ubiquitous low frequency $\omega$ fluctuations 
are physically generated in magnetized turbulence, including all the compressible modes
via the mode analysis \cite{CL03} in the global frame of reference\cite{2000ApJ...539..273C}. 
We propose a new broadened Lorentzian profile that can fit the simulation results. 
This profile allows us to quantify the contributions by different mode groups, 
enabling a better understanding of the frequency properties.We also discuss the implications of this new model for the temporal behavior of MHD turbulence.

\section*{A broadened Lorentzian model}

In the linear wave theory \cite{2002PhRvL..88x5001C,CL03}, 
small amplitude turbulence fluctuations can be viewed as simple harmonic oscillators with 
natural frequencies corresponding to the response frequencies of the three MHD {\it waves}
(see Eq. \ref{eq:wavespeed}). The nonlinear terms (${\bf v}\cdot \nabla {\bf v}$, 
${\bf \delta B}\cdot \nabla {\bf \delta B}$) can be modeled as damping terms in the 
equation of motion \cite{2009PhPl...16f2304D}. 
In this scenario, the turbulence system can be seen as a collection of damped 
harmonic oscillators (resembling an oscillating Langevin antenna in the 
case of incompressible Alfv\'enic turbulence \cite{2014CoPhC.185..578T}). 
By modeling the nonlinear term in the equation of motion as $\omega_{nl}{\bf v}$, 
with the exact functional form of $\omega_{nl}({\bf k})$ to be discussed in detail, 
the spatial-temporal energy distribution function for a selected global frame wavevector 
${\bf k}=(k_\parallel,k_\perp)$ and a MHD mode is a broadened Lorentzian distribution
(see Supplementary material for deviation):
\begin{equation}
E({\bf k},\omega) = \int dt e^{i\omega t} E({\bf k},t) \propto \frac{\omega_{nl}^2\omega_{\text{wave}}}{(\omega^2-\omega_{\text{wave}}^2)^2 + \omega_{nl}^2 (\omega + v_A |{\bf k}|\mu)^2}
\label{eq:Lorenzian}
\end{equation}
where $\omega_{\text{wave}}$ is the wave frequency of the individual mode 
(see Eq. \ref{eq:wavespeed}), $v_A$ is the Alfv\'en speed, and 
$\mu = \hat{\bf k}\cdot \hat{\bf B}$. Consequently, the dispersion relations of the 
three MHD modes no longer follow the linear form (Eq. \ref{eq:wavespeed}), 
but are modulated by the nonlinear term (see Supplementary Material). 

\begin{figure}
\includegraphics[width=1.05\textwidth]{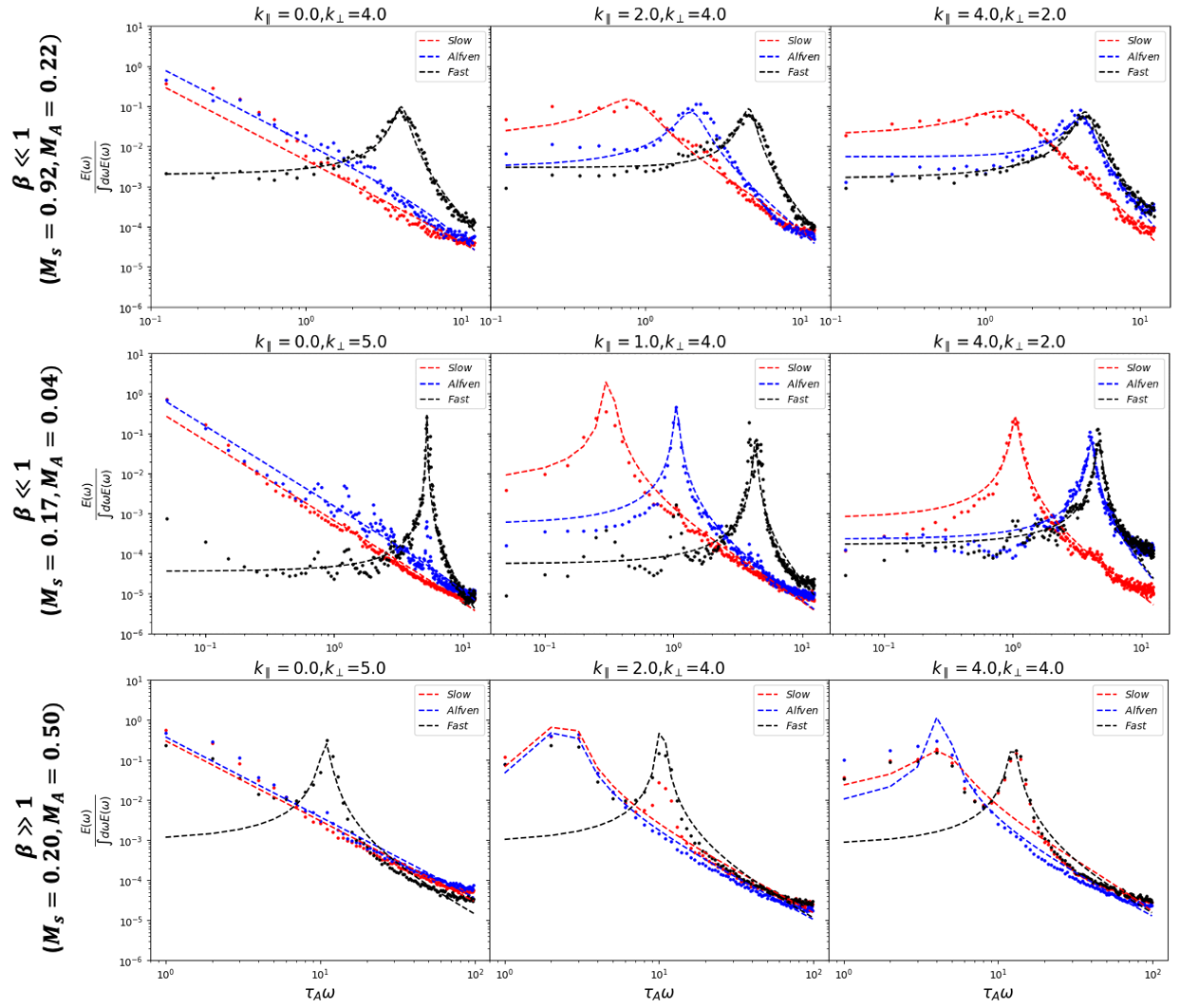}
\caption{  Temporal power $E(\omega)$ vs $\omega$ of velocity fluctuations showing how MHD turbulence with different $\beta$ (upper row: $\beta \ll 1$, simulation A1; middle row $\beta \ll 1$, A2; lower row, $\beta \gg 1$, simulation A0; Blue: Alfv\'en, Red: Slow and Black: Fast mode, see Table \ref{tab:sim}) produces significant fraction of low frequency fluctuations from Lorentzian broadening (the scattered points in each panel, c.f. Eq.\ref{eq:lorentzian_dho}) for three different regimes of $k$. From the left: $k_\parallel =0$, $k_\parallel < k_\perp$, $k_\parallel \ge k_\perp$. Each curve is normalized by its own fluctuation power for a particular choice of wavenumber $(k_\parallel, k_\perp)$ and fitted with Eq.\ref{eq:Lorenzian} (dash lines in each panel) . The x-axis of each panel is normalized with respect to the Alfv\'enic frequency  ($\tau_A^{-1}$) of the corresponding simulation. }
\label{fig:Ew}
\end{figure}

Fig. \ref{fig:Ew} shows the $E(\omega)-\omega$ diagram, each curve {\it normalized} by its own spectral power for a given wavevector (See Eq.\ref{eq:E_fig1}).
Three MHD modes at selected values of ${\bf k} = (k_\parallel, k_\perp)$ 
{  and different plasma $\beta$ (ratio of thermal to magnetic pressure, see Tab.\ref{tab:symbols} for the definition of symbols) are plotted}, 
where the fluctuations are separated according to the mode decomposition 
algorithm\cite{CL03}. The mode decomposition algorithm assumes negligible contributions from the nonlinear terms, which is not true particularly when $k_\parallel$ is small, decreasing the accuracy of the mode classifications (See Supplementary Material for discussions of the caveat of the mode decomposition method in the strongly nonlinear systems). 
{  In Fig.\ref{fig:Ew}, we intentionally include two cases for the same 
$\beta$ to illustrate that higher turbulence levels (indicated by higher sonic Mach number $M_s$ and Alfv\'enic Mach number $M_A$) produce more broadened Lorentzians (First two row of Fig.\ref{fig:Ew}). All $E(\omega)$ curves are fitted by Eq.\ref{eq:Lorenzian}
where we have made $\omega_{nl}({\bf k})$ a fitting variable for different
wave modes.} Furthermore, the peaks of the Lorentzian profiles correspond to the {    wave eigenfrequency}
of different wave modes calculated from plasma $\beta$
and wavevector ${\bf k}$ (Eq.\ref{eq:cho2}). The broadening behavior is generally consistent with previous literature\cite{2009PhPl...16f2304D,2012PhPl...19e5901T,
2014CoPhC.185..578T} and numerical simulations 
\cite{2021ApJ...922..240B,4DFFT,2022ApJ...936..127F}. 
The nonlinear broadening by $\omega_{nl}({\bf k})$ for each peak has 
a strong effect, extending the frequency distribution to both very low and high frequency limits. Notice that the analysis is performed in the global frame, 
how the local frame fluctuations 
are mapped into the global frame fluctuations measured both numerically \cite{2007PhRvE..76c6305D,2009PhPl...16f2304D,2019MNRAS.488..859Y,4DFFT} 
and observationally \cite{2023ApJ...944...98Z,2023arXiv230106709Z} 
will be addressed in the later section.

\begin{figure*}
\centering
\includegraphics[width=0.64\textwidth,height=0.40\textwidth]{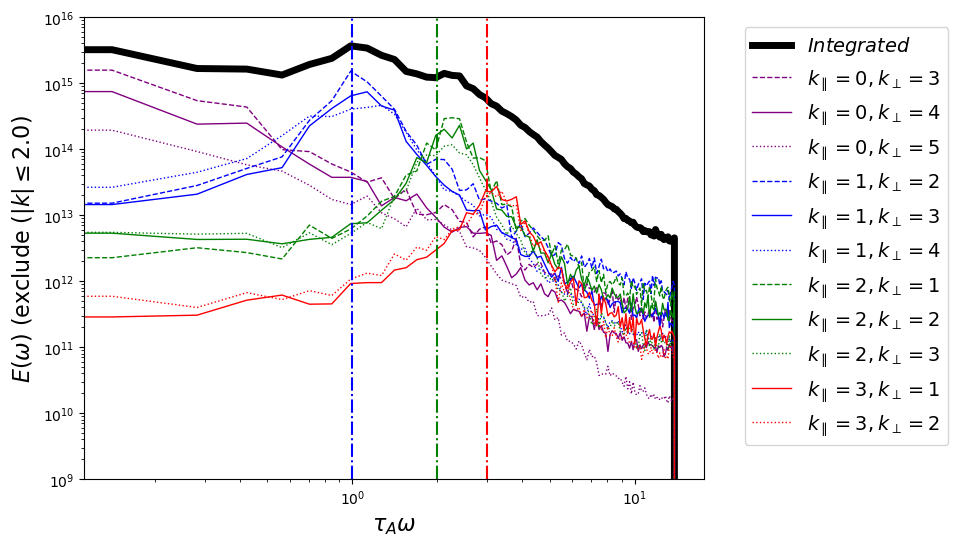}
\caption{   $E(\omega)-\omega$ curves for the 
frequency power spectra of the run A1 (See Tab. \ref{tab:sim}) where only Alfv\'enic modes are retained for different choices of ${\bf k}$. The vertical dashed lines denote the wave eigen-frequencies. 
}
\label{fig:lowomega}
\end{figure*}

\section*{Non-zero low frequency fluctuations are produced by {  nonlinear} interactions}

\begin{figure}[h]
\centering
\includegraphics[width=1.0\textwidth]{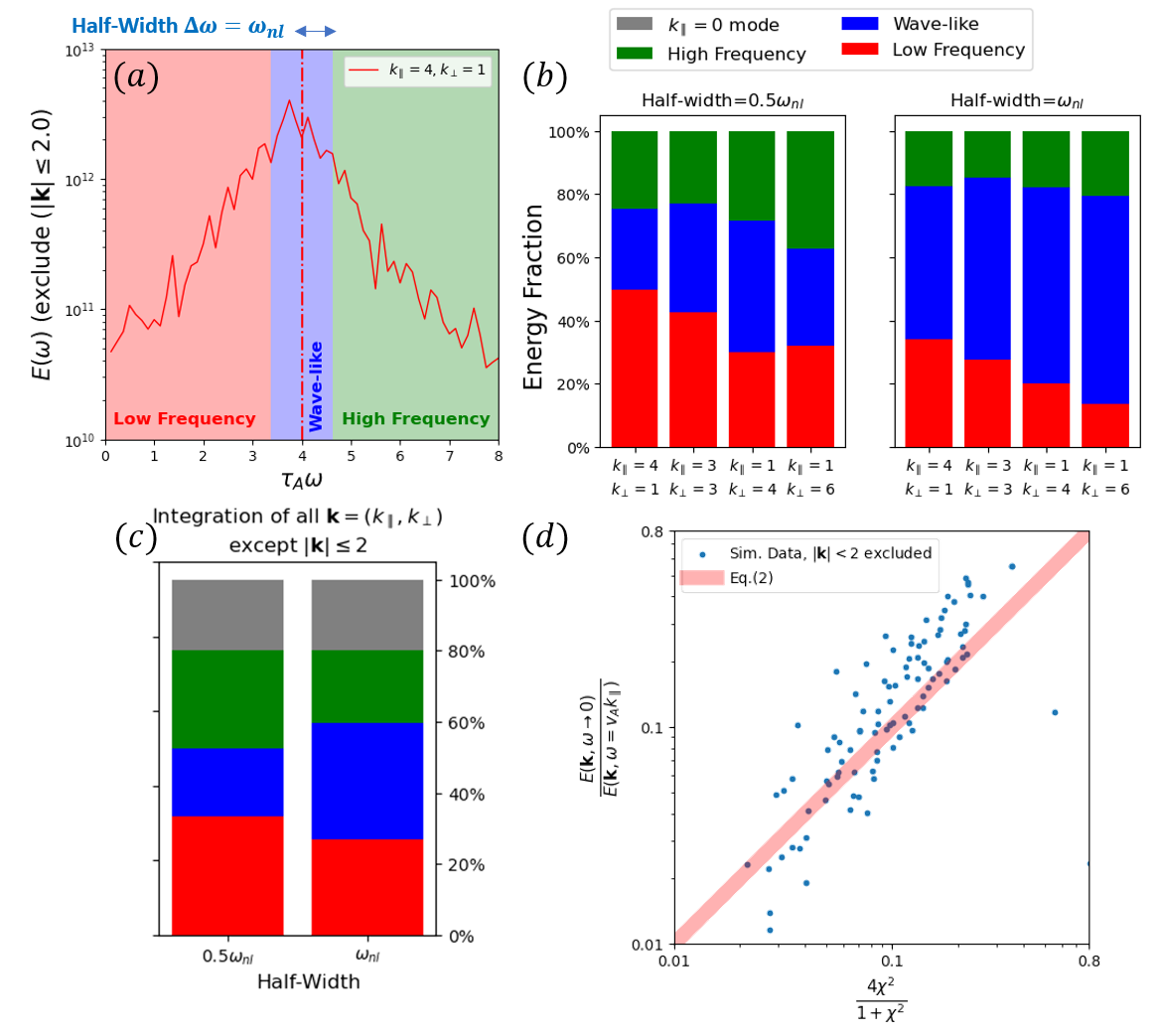}
\caption{(a) The definition of low and high frequency fluctuations with respect to the wave peak given a particular half-width. In this example the half-width is $\omega_{nl}$. (b) Relative energy fraction of Alfv\'en modes classified by frequencies in simulation A1. (c) Relative energy fraction for integrated temporal power, where modes with global $k_\parallel=0$ are also considered. (d) A comparison between the measured $\frac{E({\bf k},\omega \rightarrow 0)}{E({\bf k},\omega =v_A k_\parallel)}$ and the theoretically predicted value. The red line denotes the equality in Eq.\ref{eq:Lorenzian_w0}. }
\label{fig:fraction}
\end{figure}

Fig.\ref{fig:Ew} also highlights another important aspect, namely the origin of the low $\omega$ fluctuations.
For instance, it can be seen that the $k_\parallel = 0$ modes in all cases have an increasing power towards low $\omega$.
To quantity their contributions more clearly, Fig. \ref{fig:lowomega} shows the 
frequency power spectra of the run A1 (See Tab.\ref{tab:sim}) where we extract only the Alfv\'en mode powers. 
For frequency significantly less than $\tau_{A}^{-1} = v_A/L_{box}$ , there are two types of contributions: those with $k_\parallel > 0$ and those with $k_\parallel = 0$.
Note that we have excluded the contributions by the ${\bf k}$ modes within the injection region.
For the  $k_\parallel > 0$ modes, their contribution at low frequency, e.g., $\tau_{A}\omega = 0.1$, is very small,
and they are mainly from the low frequency wing of the Lorentzian broadened fluctuations.
Modes with $k_\parallel = 0$  (e.g., $k_\perp = 3, 4$) dominate the low $\omega$ power.
This implies that these finite frequency fluctuations above $\omega=0$ are a result of the nonlinear interaction since their Alfvén wave frequencies are zero when $k_\parallel=0$. 
The broadening of $E(\omega)$ at $k_\parallel = 0$ is always significant as long 
as $\omega_{nl}$ is non-zero. 
The non-stationary nature (i.e. $E(\omega)\neq 0$ when $\omega\neq 0$) of the $k_\parallel=0$ mode is consistent with the earlier literature that suggests the nonlinear 
terms (${v}\cdot \nabla {v}$, ${\delta B}\cdot \nabla {\delta B}$) require the 
$k_\parallel=0$ mode for efficient energy transfer and the purely perpendicular 2D cascade
\cite{1995ApJ...447..706M,GS95,1996ApJ...465..845N,2009PhPl...16f2304D}.

For finite $k_\parallel$ Alfv\'enic fluctuations, one can further quantify the ratio of 
nonlinear component versus the linear (wave) component. Using the incompressible 
Alfv\'enic fluctuations as an example,  the analytical model (Eq. \ref{eq:Lorenzian}) 
fits the simulations very well. 
For the ease of discussion, we will define the wave-like component as the integrated 
power around the linear wave frequency $\omega_{\rm wave}({\bf k})$  
within $\pm \Delta \omega ({\bf k})$, where
$\Delta \omega ({\bf k})$ is the half-width. 
We further denote the fluctuations below $\omega_{\rm wave}-\Delta \omega$ 
and above $\omega > \omega_{\rm wave} + \Delta \omega$  as {\it low} and {\it high}
frequency fluctuations, respectively (Panel (a) of Fig.\ref{fig:fraction}). 
For quantitative analysis, we consider two choices
of $\Delta \omega ({\bf k}) = 0.5$ and $1$ $\omega_{nl}({\bf k})$.
Using simulation A1, we present the relative fraction of the
wave-like, low and high frequency fluctuations for different combinations of non-zero
$(k_\parallel, k_\perp)$ in Panel (b) of Fig.\ref{fig:fraction}. 
It can be seen that significant fraction resides in the low and high frequency ranges, 
and the wave-like fraction increases when using the larger $\Delta \omega$.
Integrating over all ${\bf k}$ modes (including $k_\parallel = 0$ but
excluding the injection scale), we plot the relative contributions from
the low, wave-like, high, and $k_\parallel = 0$ components, respectively,  in Panel (c), 
again for the two choices of $\Delta \omega$. Note that the relative fraction
of the $k_\parallel = 0$ modes is a strong function of the minimum frequency 
used in the analysis
(which is chosen to be $0.1 \omega_{A}$ in this plot). 
The fraction contributed by the $k_\parallel = 0$ modes is expected to increase if a
smaller minimum $\omega$, i.e. a longer time series, is employed in the analysis.

The strength of the Lorentzian broadening is determined by the ratio of 
$\omega_{nl} \omega_{\cal{A}}$ and $\omega^2_{wave}$ (see Eq. \ref{eq:Lorenzian}). 
When $k_\parallel L_{inj} \gg 1$, 
the wave propagation frequency dominates over the nonlinear feature, resulting in 
a sharper peak in the $E(\omega)$ distribution (e.g. middle row of Fig. \ref{fig:Ew}). 
Meanwhile, $E(\omega\rightarrow 0)$ is a non-zero constant when $k_\parallel>0$, giving
\begin{equation}
\frac{E({\bf k},\omega \rightarrow 0)}{E({\bf k},\omega =v_A k_\parallel)} = 
\frac{4\omega_{nl}^2 / \omega_{\cal{A}}^2} {1 + \omega_{nl}^2 / \omega_{\cal{A}}^2}  = \frac{4\chi^2}{1+\chi^2}~~,
\label{eq:Lorenzian_w0}
\end{equation}
where $\chi = \omega_{nl}({\bf k})/\omega_{\cal{A}}({\bf k})$. Notice that 
Eq.\ref{eq:Lorenzian_w0} can be larger than 1 when $\omega_{nl} \gg \omega_{wave}$, 
indicating that the low frequency fluctuations could have a higher amplitude even 
compared to that at wave eigenfrequencies.
To verify Eq.\ref{eq:Lorenzian_w0}, we compute the numerical value of 
$E({\bf k},\omega \rightarrow 0)/E({\bf k},\omega =v_A k_\parallel)$ in simulation A1 
and compare them to the predicted values using Eq.(2) in Panel (d) of Fig.\ref{fig:fraction},
where we have used the $E(\omega)$ curves from simulation A1 with $2<|{\bf k}|<10$.  
We observe a reasonable agreement between the numerical data to the theoretical prediction 
(Eq.\ref{eq:Lorenzian_w0}), though some data points are $\sim 2$ larger than the theoretical values.
This deviation is mainly due to the fact that the mode decomposition \cite{CL03} is increasingly 
inaccurate as $\chi$ increases.

\section*{Lorentzian broadening of the compressible modes and their low-frequency contributions}

\begin{figure}[h]
\includegraphics[width=0.99\textwidth]{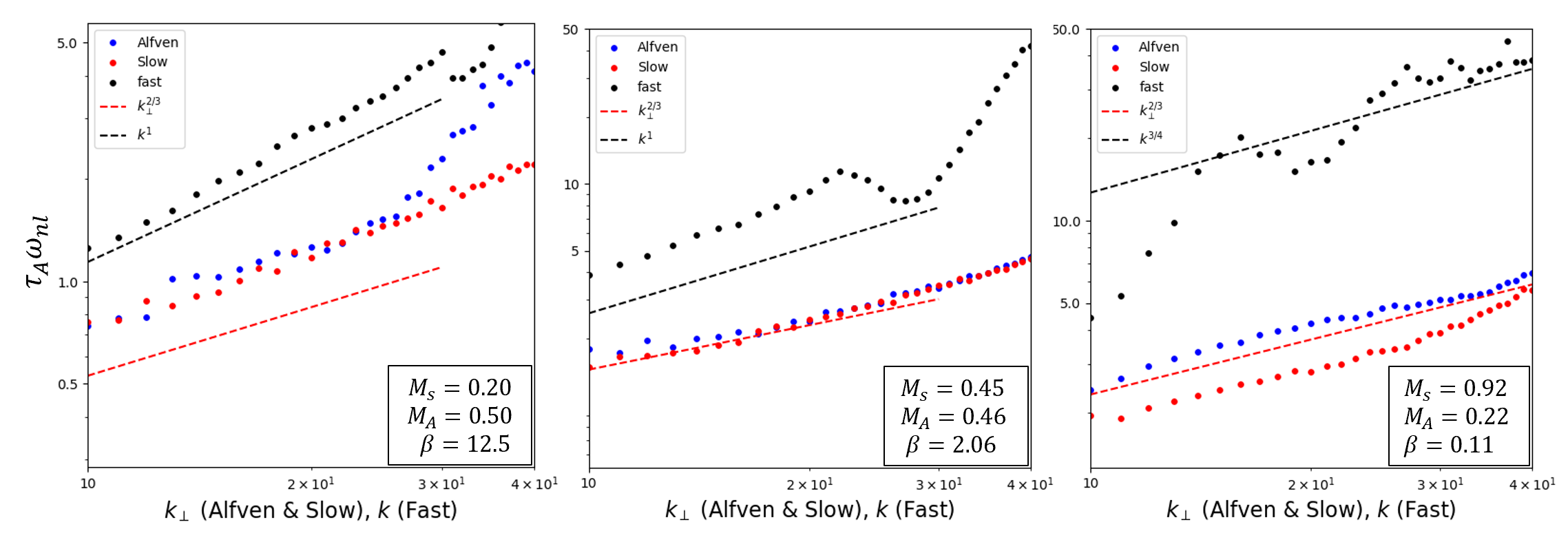}
\caption{  $\omega_{nl}$ (in units of $\tau_A^{-1}$) vs. $k_\perp$ (with $k_\parallel=3$) or $|{\bf k}|$ for three different modes (Blue: Alfv\'en, Red: Slow and Black: Fast mode). The red dash line in each panel corresponds to $\omega_{nl}\propto k_\perp^{2/3}$ while the black line corresponds to $\omega_{nl}\propto k^{1}$ for the left two panels, and $k^{3/4}$ for the right panel.). The calculations are based on runs A0, A3 and A1, respectively.   }
\label{fig:scaling}
\end{figure}

The Lorentzian broadening exists in all three MHD modes but {  the broadening strength and behavior are different} (Fig .\ref{fig:Ew}). For the case of $k_\parallel=0$ modes, the low frequency fluctuations are mostly contributed by both Alfv\'en and slow modes, where the exact ratio is determined by the relative energy fraction of the two modes (left column of Fig. \ref{fig:Ew}). 
As discussed previously, the broadening width at $k_\parallel=0$ only depends on the value of $\omega_{nl}$. 
The similar slopes in $\omega$ for both Alfv\'en and slow modes' temporal power spectrum $E(\omega)$ suggest that their $\omega_{nl}$ is similar in magnitude (See also Fig.\ref{fig:scaling}). 

For modes with $k_\parallel > 0$, slow mode contributes more low-frequency fluctuations than the other two modes due to its lower $\omega_{wave}$ (middle and right columns of Fig. \ref{fig:Ew}). 
This behavior is more amplified in low $\beta$ where the slow wave speed is significantly smaller than the Alfv\'en speed for the same ${\bf k}$. 
Different from the case of $k_\parallel=0$, the relative fraction of low frequency fluctuations for $k_\parallel>0$ is governed by $\chi$ parameter (Eq.\ref{eq:Lorenzian_w0}), which is the largest for slow modes. 
Therefore the Lorentzian broadening is effectively stronger for slow modes,  albeit the mode fractions between slow and other two modes have to be taken into account\cite{2020PhRvX..10c1021M}. In contrast, fast mode plays a negligible role in low-frequency fluctuations because its wave speed is significantly higher than those of Alfv\'en and slow modes, and is always non-zero unless $|{\bf k}|=0$.

\section*{Scalings of nonlinear frequency $\omega_{nl}$ for different modes}

To quantify the nonlinear broadening for each MHD mode and their dependence on turbulence properties, we can use Eq. \ref{eq:Lorenzian} to extract $\omega_{nl}$ from the Lorentzian profiles. 
Fig. \ref{fig:scaling} shows the nonlinear frequencies of three subsonic, 
sub-Alfv\'enic simulations with various $\beta$ as a function of $k_\perp$ or $k=|{\bf k}|$. 
The effect of the local and global reference frame \cite{leakage} is {\it less} significant when applying the mode decomposition technique \cite{CL03} since $M_A$ is small for our simulation (Tab.\ref{tab:sim}). 
In the case of high $\beta$ (Left panel of Fig. \ref{fig:scaling}), both Alfv\'en and slow modes have their $\omega_{nl}$ scale in the same way, $\propto k_\perp^{2/3}$. 
Fast modes cascade isotropically \cite{CL03}, and therefore we plot the $\omega_{nl}$ for fast modes along $k$. 
We find a scaling of $\omega_{nl}\propto k^{1}$ best fit our data for high and intermediate $\beta$, suggesting a cascade of $E_k \propto k^{-2}$ for these two cases.
For low $\beta$ (Right panel of Fig.\ref{fig:scaling}), {\hy the data points are too scattered to be conclusive,} despite an apparent trend-line of $k^{3/4}$ is observed for a limited range of $k$.

How do we understand the scaling trend in Fig.\ref{fig:scaling}? It is commonly assumed in the case of strong Alfv\'enic turbulence that $\omega_{nl}\sim k_\perp \delta v_k$ \cite{GS95,LV99}.  
However, the nonlinear time $\tau_{nl}\sim\omega_{nl}^{-1}$ does not necessarily correspond to the cascade time. 
The dependence of the cascade time is actually a function of $\chi({\bf k})$, as noted in earlier literature \cite{1963AZh....40..742I,1965PhFl....8.1385K,LV99,CL03,2023JPlPh..89b9005G}. 
A rigorous closure calculation by Tripathi et al. (in prep.) shows that the 
cascade time ($\omega_{tr}$) has the following dependence on $\chi$:
\begin{equation}
\omega_{tr} \approx \frac{\omega_{nl}^2}{\omega_{\text{wave}} (1+\chi)}
\label{eq:Tripathi_Eq_2}
\end{equation}
In particular, for extreme cases of $\chi$:
\begin{equation}
\begin{aligned}
\omega_{tr} &\sim \frac{\omega_{nl}^2}{\omega_{\text{wave}}} & \quad & (\chi \ll 1) \\
\omega_{tr} &\sim \omega_{nl} & \quad & (\chi \gg 1 )
\end{aligned}
\label{eq:Tripathi_Eq_3}
\end{equation}
where the first expression is the Iroshnikov–Kraichnan 
cascade rate\cite{1963AZh....40..742I,1965PhFl....8.1385K} and the latter is commonly adopted for strong turbulence cascade \cite{GS95}.  
The constancy of the energy cascade rate $\delta v_k^2 \omega_{tr}$ allows for a quick quantification of the relation between $\omega_{nl}$ and spectral power $E_k$ as functions of $k$. Writing:
\begin{equation}
\text{const} \approx \begin{cases}
\frac{\delta v_k^2 \omega_{nl}^2}{\omega_{wave}} \quad (\chi \ll 1) \\
\delta v_k^2 \omega_{nl} \quad (\chi \gg 1)
\end{cases}
\end{equation}
using $E_k \sim \delta v_k^2/k$ gives:
\begin{equation}
E_k \propto \begin{cases}
\omega_{nl}^{-2} \quad (\chi \ll 1) \\
k^{-1}\omega_{nl}^{-1} \quad (\chi \gg 1)
\end{cases}
\label{eq:HR_E_k_omega}
\end{equation}

In the incompressible Alfv\'en and high-$\beta$ slow mode limit, where $E_k \propto k_\perp^{-5/3}$, which implies that $\omega_{nl}\propto k_\perp^{2/3}$, which are observed in Fig.\ref{fig:scaling} for both Alfv\'en and slow modes in all choices of $\beta$. 
For fast modes, the Iroshnikov–Kraichnan spectrum \cite{1963AZh....40..742I,1965PhFl....8.1385K,2023JPlPh..89b9005G} 
($E_k\propto k^{-3/2}$) suggests $\omega_{nl}\propto k^{3/4}$, which is only observed in the case of low $\beta$ case.
For the measured scaling $\omega_{nl}\propto k^{1}$ in the left and middle panels of Fig. \ref{fig:scaling}, Eq. \ref{eq:HR_E_k_omega} gives $E_k\propto k^{-2}$, which is commonly proposed as the alternative scaling of fast modes \cite{2020PhRvX..10c1021M}. 

\section*{Discussion}

The Lorentzian broadening effect is intrinsically caused by the nonlinear effects in turbulence. 
However, the results are obtained in the global frame with respect to the mean background
magnetic field direction. 
In the case of incompressible Alfv\'enic turbulence, many MHD turbulence 
studies \cite{GS95,LV99,2000PhRvL..85.4656C} have emphasized that the nonlinear 
fluctuations are generated perpendicular to the local magnetic fields, characterized
by the local wavevector $(k_{\parallel,L}, k_{\perp,L})$. 
These fluctuations, when viewed in a global frame, could lead to a 
{\it broadening} effect in $\omega$. 
This can be understood using the following picture:
the spectral power of a 
particular global wavevector $(k_\parallel,k_\perp)$ measured in the global frame
could collect many eddies of different sizes in the local frame
\cite{2000ApJ...539..273C}.  
These eddies will contribute different spectral weights, 
with particularly higher weights for the local eddies that have 
dimensions  $(k_{\parallel,L}^{-1}, k_{\perp,L}^{-1})$ close to $(k_{\parallel}^{-1},k_{\perp}^{-1})$. 
Each of these local wavevectors is projected onto the global 
$k_\parallel$ axis, which gives a measured $\omega$ as 
$\sim k_{\parallel,L} v_A$, along with its contribution to the $E(\omega)$ spectrum.
The variation of $k_\parallel$ due to the projections 
of the local wavevectors is then translated to the $\omega$ space, 
resulting in a broadening of the $E(\omega)$ spectrum.

The broadening width can be approximated as follows: 
The wavevector difference between the local and global frames 
in the case of Alfv\'enic turbulence is strongly correlated to the 
Alfv\'enic Mach number $M_A$ \cite{leakage}, i.e., 
$\delta k_\parallel/k_\parallel \sim M_A \sim \delta v/v_A$.  
As a result, the average dispersion of $\omega$ is 
$\delta \omega \sim \delta k_\parallel v_A \sim k_\parallel \delta v$, 
centered at the global frame $\omega \sim k_\parallel v_A$. 
In other words, the frame transformation naturally generates a 
broadening in $\omega$ as long as $M_A$ is non-zero.
In comparison, the dispersion relation of fast modes has only weak 
dependence on the wave vector direction. 
The projection effect for fast modes is therefore marginal, 
accounting partially for their weaker nonlinear behavior.

As a remark, the concept of critical balance\cite{GS95}
is also closely connected to the nature of non-zero low-frequency fluctuations.
It has been argued that the nonlinear timescale of incompressible 
Alfvén mode is approximately equal to its wave propagation timescale, 
i.e., $\chi=1$, forming the basis of modern MHD turbulence theory 
\cite{GS95,CL03}, despite dissent views remain \cite{2016PhRvL.116j5002M,2020ApJ...897...37O,2020PhPl...27f2308C}. 
Refinement of the critical balance theory has been proposed in the 
community in the incompressible limit \cite{2015MNRAS.449L..77M}.
Understanding the low-frequency fluctuations provides valuable insights 
into the nonlinear timescales, as there is a designated anisotropy 
scaling ($k_\parallel\propto k_\perp^{2/3}$ in the case of Alfv\'en modes) 
related to the critical balance. 
However, whether such balance exists for the compressible modes 
is still an unresolved question \cite{2023JPlPh..89b9005G}.

\clearpage

\section*{Data Availability}

The data used in this work are listed in Tab.\ref{tab:sim}. Data and the input file will be available upon request.
\begin{table*}[t]
\caption{Table of simulations used in the current work \label{tab:sim}}
\small
\centering
\begin{tabular}{ccccc}
\hline
\hline
Model Name 
& $M_s$ & $M_A$ & $\beta$ &Energy Injection Rate  \\ 
\hline\hline 
A0 & 0.20 & 0.50 & 12.5 &  0.01     \\
A1 & 0.92 & 0.22 & 0.11 & 0.01      \\
A2 & 0.17 & 0.04 & 0.13 & 0.0001   \\
A3 & 0.45 & 0.46 & 2.06 & 0.001    \\
\hline\hline
\multicolumn{5}{l}{The default parameters are: }\\
\multicolumn{5}{l}{$\quad$ $c_s = 1$, $L_{box}=1$, $L_{inj}\ge 1/2$, $\langle\rho\rangle=1$,}\\
\multicolumn{5}{l}{ $\quad$  Injection frequency $\tau_s\omega_{inj}=100$, resolution = $512^3$. }\\
\end{tabular}
\end{table*}

\begin{table*}[t]
\caption{Table of symbols used in this paper\label{tab:symbols}}
\small
\centering
\begin{tabular}{cc}
\hline
\hline
Symbol & Meaning  \\
\hline
\multicolumn{2}{l}{\bf Shorthand Operators}\\
$\langle x\rangle$ & Spatial average of x.\\
$\hat{\bf x}$ & Unit vector of ${\bf x}$.\\
$\tilde{x}$ & Spatial-temporal Fourier transform of x.\\\hline
\multicolumn{2}{l}{\bf Plasma Parameters}\\
$\rho$ & Fluid density.\\
${\bf v}$ & Fluid velocity vector, R.M.S. value of it =$v_{turb}$.\\
${\bf B}_0$ & Mean Magnetic Field vector, its unit vector is sometimes denoted as $\hat{\lambda}$.\\
${\bf \delta B}$ & $={\bf B}-{\bf B}_0$, the mean-subtracted magnetic field.\\
$v_{inj}$ & Injection velocity amplitude.\\
$v_A$ & $=\frac{|{\bf B_0}|}{\sqrt{4\pi\langle \rho\rangle}}$, Alfv\'enic Speed. The vectorized form: ${\bf v}_A=\frac{\bf {\bf B}_0 }{\sqrt{4\pi\langle \rho\rangle}}$\\
$c_S$ & Sonic Speed. \\
$M_s$ & $= v_{inj}/c_s$, Sonic Mach number.\\
$M_A$ & $= v_{inj}/v_A$, Alfv\'enic Mach number.\\
$\beta$ & $=2M_A^2/M_s^2$, plasma compressibility.\\
$L_{box}$ & Simulation domain size, always $=1$ in our set-up. \\
$L_{inj}$ & Injection scale, $L_{box}/2$.\\
\hline
\multicolumn{2}{l}{\bf Fourier space \& Mode parameters}\\
$E(\omega)$ & = $\tilde{E}(t)$, temporal energy spectrum.\\
$\omega$  & The temporal frequency.\\
${\bf k}$ & Wavevector (3D vector). \\
$\delta v_k$ & Spatial Fourier amplitude of $v$ at ${\bf k}$.\\
$k_\parallel, k_\perp$ & The parallel and perpendicular wavenumber in the global frame of reference.\\
$\mu=\cos\theta $ & $=\hat{\bf k}\cdot \hat{\bf B_0}$, the cosine of the angle $\theta$ between wavevector and mean magnetic field.\\
${\cal A},{\cal S},{\cal F}$ & Alfv\'en, Slow, Fast Mode subscripts.\\
$\alpha$ & $ = \beta/2$.\\
$\cal{D}$ & $=(1+\alpha)^2- 4\alpha\cos ^2\theta$.\\
$\zeta_{{\cal A},{\cal S},{\cal F}}$ & Linearized Alfv\'enic, Slow and Fast wave vectors (c.f. Eq.\ref{eq:cho2}).\\
\hline
\multicolumn{2}{l}{\bf Timescales and nonlinearity parameters}\\
$\tau_s,\tau_A$ & = $L_{box}/c_s,L_{box}/v_A$, the sound crossing time and Alfv\'enic time, respectively.\\ 
$\omega_{nl}$ & Nonlinear frequency.\\
$\omega_{tr}$ & Cascade frequency, c.f. Eq.\ref{eq:Tripathi_Eq_2}.\\
$\omega_{wave}$ & Wave frequency, c.f .Eq.\ref{eq:wavespeed}.\\
$\omega_{inj}$ & Injection Frequency.\\
$\chi $ & Natural nonlinear parameter, defined as $\omega_{nl}/\omega_{wave}$.\\
\hline\hline
\end{tabular}
\end{table*}

\section*{Code Availability}
\subsection*{Numerical simulations}
The numerical simulations are performed with Athena++ \cite{2020ApJS..249....4S}.We summarize the simulations in Tab. \ref{tab:sim}.  Our data are time series of three-dimensional, triply periodic, isothermal MHD simulations with continuous force driving via {\it direct spectral injection} \footnote{The reason why we did not employ the Ornstein–Uhlenbeck forcing is because we would like to have more control on the injected values of ${\bf k}$ and the injection frequency $\omega_{inj}$.} unless specified. We run our simulations for at least 10 sound crossing time ($\tau_s=L_{box}/c_s$) and take snapshots at $\Delta \tau = \tau_s/100$ to ensure the time-axis sampling satisfies the condition that:
\begin{equation}
\Delta \tau_{required} < \frac{L_{box}}{v_{fastest}}
\label{eq:time_dur_req}
\end{equation}
where $L_{box}$ is the size of the simulation domain, and $v_{fastest}$ is the fastest speed in the numerical simulations. The typical parameters of our simulations are listed in Tab.\ref{tab:sim}. The injection is performed so that we only have eddies at scales $L_{inj}/L_{box}\ge 1/2$, which corresponds to $|{\bf k}|\le 2$. All simulations are driven solendoially.  All numerical simulations are truncated in Fourier space to $128^3$ regardless of its original size to save computational resources. That will not change the statistics of spatio-temporal spectrum for $|{\bf k}|<128$.

\subsection*{Analysis}

Analysis are performed by {\bf \it Julia} with the packages available upon request.

\bibliography{refs.bib} 
\clearpage

\section*{Competing interests}
{   The authors declare no competing interests.}

\section*{Supplementary information}

\section*{Acknowledgements}
Research presented in this article was supported by the Laboratory Directed Research and Development program of Los Alamos National Laboratory under project numbers 20220700PRD1 and 20220107DR, and a U.S. Department of Energy Fusion Energy Science project. This research used resources provided by the Los Alamos National Laboratory Institutional Computing Program, which is supported by the U.S. Department of Energy National Nuclear Security Administration under Contract No. 89233218CNA000001. This research also used resources of the National Energy Research Scientific Computing Center (NERSC), a U.S. Department of Energy Office of Science User Facility located at Lawrence Berkeley National Laboratory, operated under Contract No. DE-AC02-05CH11231 using NERSC award HEP-ERCAP-m2407,m3122 (PI: Fan Guo, LANL) and FES-ERCAP-m4239 (PI: KHY, LANL). {Inspiring discussions with Jungyeon Cho (Chungnam National University), Patrick Diamond (UCSD), Fan Guo (LANL), Alex Lazarian (UW Madison),  Willian Matthaeus (U. Delaware), Bindesh Tripathi (UW Madison), Ka Wai Ho (UW Madison/LANL) and Ethan Vishniac (JHU)} are acknowledged.

\section*{Author contributions}
Yuen and Li initiated and designed the project. Yuen performed the simulations and analyzed the main result of the manuscript. Yuen and Yan derived the nonlinear timescales. All authors contributed in writing, editing and approving the manuscript.

\section*{Methodology}
\label{sec:methodology}

The simulations shown in Tab.\ref{tab:sim} are in the form of time series, say ${\bf v}({\bf r},t)$. We perform a full 4D Fourier transform:
\begin{equation}
{\bf v}({\bf r},t) = \sum_{{\bf k},\omega} \tilde{\bf v}({\bf k},\omega) e^{i({\bf k}\cdot {\bf r}-\omega t)}
\end{equation}
Taking the cylindrical coordinate ${\bf k} \rightarrow (k_\parallel,k_\perp)$, the spatio-temporal power spectrum $E(k_\parallel,k_\perp,\omega)$ is given by:
\begin{equation}
E(k_\parallel,k_\perp,\omega) = \oint k_\perp d\Omega  |\tilde{\bf v}({\bf k},\omega)|^2 = 2\pi k_\perp |\tilde{\bf v}({\bf k}=(k_\parallel,k_\perp),\omega)|^2 
\label{eq:9}
\end{equation}
which cautions have to be taken that the coordinate transform adds an additional factor of $k_\perp$. 

\subsection*{Mode Decomposition when $\chi \ll 1$}

The MHD modes are obtained by performing the dot product $\tilde{\bf v}_X = (\tilde{\bf v}\cdot \hat{\zeta}_X) \hat{\zeta}_X$ to the turbulence variables, where $X=\cal{A}\text{(Alfv\'en)},\cal{S}\text{(Slow)},\cal{F}\text{(Fast)}$ and $\zeta_X$ are\cite{CL03}:
\begin{equation}
    \begin{aligned}
    \zeta_{\cal A} &\propto \hat{\bf k} \times \hat{\bf  \lambda}\\
    \zeta_{\cal S} &\propto (-1 +\alpha-\sqrt{\cal{D}}) ({\bf k}\cdot \hat{\bf \lambda}) \hat{\bf \lambda}  + (1+\alpha - \sqrt{\cal{D}}) ( \hat{\bf \lambda} \times ({\bf k}\times \hat{\bf \lambda})) \\
    \zeta_{\cal F} &\propto (-1 +\alpha+\sqrt{\cal{D}}) ({\bf k}\cdot \hat{\bf \lambda}) \hat{\bf \lambda}  + (1+\alpha + \sqrt{\cal{D}}) ( \hat{\bf \lambda} \times ({\bf k}\times \hat{\bf \lambda})) \\
    \end{aligned}
    \label{eq:cho2}
\end{equation}
where $\lambda = \langle \hat{\bf B}\rangle$, $\alpha = \beta/2$, $\beta=2M_A^2/M_s^2$, ${\cal{D}}=(1+\alpha)^2- 4\alpha\cos ^2\theta$, $ \cos\theta =\mu = \hat{\bf k}\cdot \hat{\bf \lambda}$.  We have to emphasize that the decomposition was proposed assuming negligible non-linear terms \cite{CL03}, as they are treated as {\it second} order terms and should be zeroed out after linear perturbations. Furthermore, the decomposition is performed in the global frame of reference\cite{2000ApJ...539..273C}. For the case where $\chi$ is not small, we expect the decomposition to be less accurate. Example signatures include the lower right panel of Fig.\ref{fig:Ew}, where fast wave eigen-peak appears in slow mode $E(\omega)$. However, a simple eigenvalue analysis suggests that the appearance of the non-linear term {\it does not} change the {\it mean} direction of the eigenvectors to the {  second} order.

The 4D energy spectrum of a particular mode is given by (c.f. Eq.\ref{eq:9}):
\begin{equation}
E_{X=({\cal A},{\cal S},{\cal F})}(k_\parallel,k_\perp,\omega) = \oint k_\perp d\Omega |\tilde{\bf v}({\bf k},\omega)|^2 = 2\pi k_\perp |\tilde{\bf v}({\bf k}=(k_\parallel,k_\perp),\omega)\cdot \hat{\zeta}_X|^2 
\end{equation}

\subsection*{    The quantitative model of MHD turbulence}
{ 

\subsubsection*{Deviation of the Lorentzian function (Eq.\ref{eq:Lorenzian}) from MHD equations}
To derive the Lorentzian function used in Eq.\ref{eq:Lorenzian}, we need to make the following approximations to the ideal MHD equations with isothermal equation of state.
\begin{enumerate}
    \item We assume first order perturbations on $\rho, v$ \& $B$, ${\bf B} = {\bf B}_0 + \delta {\bf B}, z = ln(\rho/\langle \rho \rangle), $ and write the Alfv\'enic vector ${\bf v}_A=\frac{{\bf B}_0}{\sqrt{4\pi\langle \rho\rangle}}$. 
    \item We preserve all nonlinear terms ($({\bf v}\cdot \nabla) {\bf v},({\bf B}\cdot \nabla) {\bf B}$) and approximate them as $\omega_{nl}{\bf v}/2$ and $\omega_{nl}{\bf B}/2$, respectively. {    The approximation is exact for pure Alfv\'enic turbulence \cite{2014CoPhC.185..578T}.}
    \item We perform Fourier transform, writing $\partial_t \rightarrow -i\omega$ and $\nabla \rightarrow i{\bf k}$. 
\end{enumerate}
Then the equation of motion in the Fourier space is given by ($c_n^2 = c_s^2 + \frac{v_A^2}{2}$):
\begin{equation}
    (-i \omega + \frac{\omega_{nl}}{2})^2 \tilde{\bf v} + {\bf k} c_n^2 {\bf k}\cdot {\bf \tilde{v}} - (i{\bf v}_A\cdot {\bf k}+\frac{\omega_{nl}}{2} )^2 \tilde{\bf v} = 0
    \label{eq:eom}
\end{equation}

The Lagrangian power spectrum is defined by the squared amplitude of the Fourier transformed velocity $\tilde{\bf v}$. Solving Eq.\ref{eq:eom} against the plane wave solution (the Fourier transform of $\delta$-function),  the Lagrangian power spectrum is given by (using $\omega_{{\cal{A},\cal{S},\cal{F}}}^2 = (v_A |{\bf k}|\mu)^2 + c_n^2 ({\bf k}\cdot {\bf \tilde{\hat{v}}})^2$, c.f. Eq.\ref{eq:wavespeed}):
\begin{equation}
E(\omega,{\bf k}) = |\tilde{\bf v}|^2 = \frac{1}{(\omega^2-\omega_{\cal{A},\cal{S},\cal{F}}^2)^2 + \omega_{nl}^2 (\omega + v_A k\mu)^2}
\label{eq:lorentzian_dho}
\end{equation}

\noindent{\bf Special cases for Incompressible regime}: In the limit of incompressible turbulence $\omega_{\cal{A}}=\omega_{\cal{S}}$ and Eq.\ref{eq:lorentzian_dho} reduces to the following form:
\begin{equation}
    E(\omega,{\bf k}) \propto \frac{1}{(\omega+\omega_{\cal A})^2} \frac{1}{(\omega-\omega_{\cal A})^2 + \omega_{nl}^2}
    \label{eq:lorentzian_incomp}
\end{equation}

\noindent{\bf Special cases for $\mu=0$}: In the limit of $\mu=0$, Eq.\ref{eq:lorentzian_dho} reduces to the following form:
\begin{equation}
    E(\omega,{\bf k}) \propto \frac{1}{\omega^2} \frac{1}{(\omega-\omega_{\cal{A},\cal{S},\cal{F}}(\mu=0))^2 + \omega_{nl}^2}
\end{equation}
\subsection*{{  Linear Wave Frequencies} and Dispersion Relations of MHD Modes}

The dispersion relations of the three MHD waves are given by\cite{CL03} ($\cal{A}=\text{Alfv\'en},\cal{S}=\text{Slow},\cal{F}=\text{Fast}$):
\begin{equation}
\begin{aligned}
\omega_{\cal{A}} &= v_A\mu |{\bf k}|  \\
\omega_{\cal{S}} &= \Big(\frac{1}{2}\left[v_A^2+c_s^2 - \sqrt{(v_A^2+c_s^2)^2 - 4v_A^2c_s^2\mu^2}\right]\Big)^{\frac{1}{2}}|{\bf k}|\\
\omega_{\cal{F}} &= \Big(\frac{1}{2}\left[v_A^2+c_s^2 + \sqrt{(v_A^2+c_s^2)^2 - 4v_A^2c_s^2\mu^2}\right]\Big)^{\frac{1}{2}}|{\bf k}|\\
\end{aligned}
\label{eq:wavespeed}
\end{equation}
where $v_A$ is the Alfv\'en speed, $c_s$ is the sound speed and $\mu = \cos\theta = \hat{\bf k}\cdot \langle \hat{\bf B}\rangle$. 
\subsection*{Lorentzian Profile}

We normalize the energy spectrum by each of its integrated value:
\begin{equation}
\bar{E}_X(\omega; k_\parallel,k_\perp) = \frac{E_X(\omega; k_\parallel,k_\perp)}{\int d\omega E_X(\omega; k_\parallel,k_\perp)}
\label{eq:E_fig1}
\end{equation}

The Lorentzian profiles (c.f. Figures \ref{fig:Ew},\ref{fig:lowomega},\ref{fig:scaling}) are fitted by least square method to the functional form 
\begin{equation}
y(\omega;[A,B,C,D])={  \frac{D}{(\omega^2-A^2)^2+(\omega+B)^2 C^2}}
\label{eq:model_lor}
\end{equation}
normalized to unity. {  Here $A\rightarrow \omega_{wave}$, $B\rightarrow {\bf v}_A \cdot {\bf k}= v_A |{\bf k}|\mu $, $C = \omega_{nl}$ (c.f. Eq.\ref{eq:Lorenzian}). } The fitting was done by a bounded 4-parameter regression on $\log(\omega)$ and $\log(1/y)$ with $A,C,D \ge 0$ and $B \in \mathcal{R}$. 

\subsection*{Scaling Relation of $\omega_{nl}$ for incompressible Alfv\'enic turbulence}

In the strong B-field limit, the nonlinear frequency for the Alfvén mode can be approximated by the Kolmogorov scaling perpendicular to the mean field (i.e., $v/v_{\text{inj}}\propto (l/L_{\text{inj}})^{1/3}$) \cite{LV99}:
\begin{equation}
\omega_{nl} = v_{\text{turb}} l_\perp^{-1} \sim v_{\text{inj}} L_{\text{inj}}^{-1/3} k_\perp^{2/3}
\label{eq:GS_ts}
\end{equation}
where $v_{\text{inj}}$ and $L_{\text{inj}}$ are the velocity and length scale at injection. {   Notice that the nonlinear time for the three different modes is actually slightly different (c.f. Fig. \ref{fig:scaling}), contrary to the common belief \cite{2023JPlPh..89b9005G}, as the nonlinear interactions act differently on the three modes. However, the conclusion that the fast modes are in weak turbulence \cite{2023JPlPh..89b9005G} when Alfvén waves are in strong turbulence \cite{GS95} in small scales is phenomenologically correct .}

\end{document}